\documentclass{ws-procs975x65}

\newcommand{\pacz}{Paczy\'nski}

\begin{document}

\title{M31 pixel lensing: Observations
at the Loiano telescope}

\author{S. Calchi Novati$^*$\\on behalf of the PLAN 
collaboration$^\flat$}

\address{Dipartimento di Fisica ``E. R. Caianiello'', 
Universit\`a di Salerno, Via Ponte don Melillo, 84081 Fisciano (SA), Italy
and INFN, sez. di Napoli, Italy\\
$^*$E-mail: novati@sa.infn.it\\
$^\flat$ http://plan.physics.unisa.it}

\begin{abstract}
We discuss pixel lensing observations towards M31 
carried out at the Loiano telescope. We have established
a fully automatic pipeline for the detection
and the characterization of microlensing events.
We have evaluated the efficiency of the pipeline. We have estimated
the expected signal by means of a Monte Carlo
simulation. As a result we select 2 microlensing candidates.
This is compatible with the expected M31 self-lensing  signal.
The small statistics of events at disposal
does not still allow us, however, to draw definite conclusions
on the content of compact halo objects.
\end{abstract}


\bodymatter

\section{Introduction}

The content of dark matter in galactic
haloes in form of compact halo objects (MACHOs)
is still an open issue\cite{massey10}. Following the original
proposal of \pacz\ \cite{pacz86}, microlensing probed
to be an efficient tool to carry out this 
research programme. The original target to explore
the MACHO content of the halo of the
Milky Way (MW) has been the Large Magellanic Cloud
(LMC). Along this line of sight the more relevant results
have been obtained by the MACHO\cite{macho00},
EROS\cite{eros07} and OGLE\cite{ogle09} collaborations.
These observational programmes allowed to exclude
MACHOs for a large range of masses as a possible
dark matter component. Still, the results
of the MACHO collaboration show that a sizeable
mass fraction of the MW halo ($f\sim~20\%$) might
be composed of compact halo objects of about $0.5~\mathrm{M}_\odot$.
This outcome has been challenged by the EROS and the OGLE
results. Once we accept that the reported candidate
microlensing events are not to be attributed
to intrinsic variable sources, a possible contamination of the 
MACHO lensing signal comes from \emph{self lensing}, namely a lensing event
where both the lens and the source belong to some
luminous population. Indeed, the fact that
the only remaining allowed mass range for MACHOs
correspond to that of the stars that might also
act as lenses may be indicative of some bias
(because the characteristics
of the microlensing events, in particular their
duration, depend from the lens mass). On the other hand,
if these events are really to be attributed
to MACHOs in this mass range, a fortiori
for such a sizeable fraction as that implied
by the MACHO collaboration results,
this might have some deeper astrophysical
meaning. The LMC microlensing,
together with microlensing towards
the Galactic center used as a probe 
to constrain the inner Galactic structure, is the
subject of the thorough review of Moniez\cite{moniez10}.

Beyond the LMC, the next suitable target for
microlensing searches is M31\cite{crotts92,agape93,jetzer94}.
It allows one to explore a different line of 
sight through the MW halo; we can fully map
the M31 own dark matter halo (which is not
possible for the MW one, this being perhaps
the most severe limitation for LMC studies);
finally, the inclination of M31 is expected
to give rise to a characteristic signature
in the spatial distribution of M31 halo events
such to facilitate their identification against
the contamination of self-lensing events.
However, because of the distance of M31, the sources
for microlensing events are no longer resolved objects
so that we enter the regime usually referred to as 
\emph{pixel lensing}\cite{gould96}. Several observational
campaigns have already been undertaken along this line of sight.
As for the issue of the dark matter compact halo object content, the 
POINT-AGAPE collaboration claimed for an evidence of a signal
in the same mass range indicated by the MACHO LMC analysis\cite{novati05}.
This outcome, however, has been challenged by
the MEGA collaboration who, using the same set of data,
concluded that their detected signal is compatible
with the expected self-lensing rate\cite{mega06}.
In fact, especially if one does not move far enough
from the M31 center, the contamination of self lensing,
still difficult to be exactly quantified,
is in any case expected to be quite large with respect to MACHO lensing.
Pixel lensing towards M31 is the subject of the review
of Calchi~Novati\cite{grg10} (we also recall
the recent analysis of Tsapras et~al.\cite{tsapras10},
where a new analysis of the POINT-AGAPE data set
is presented together with a comparison 
of the results of different pipelines). In the present contribution
we report on the ongoing pixel lensing
campaign towards M31 carried out by the PLAN collaboration.

\section{Microlensing candidate events} 

Our observational programme started in 2006 using
the 152cm Loiano telescope at the Osservatorio Astronomico di 
Bologna\footnote{\url{http://www.bo.astro.it/loiano/index.html}}.
We have been monitoring two fields of $13'\times 12.6'$
each around the M31 center (Fig.~1a) to probe both
self lensing and MACHO lensing\cite{novati07}. We observe
for consecutive (and full) nights in order
to be able to properly sample and characterize
the microlensing events and get to a
sufficient signal to noise ratio. Since 2008 we are also using
the 150cm TT1 telescope at the Osservatorio
Astronomico di Capodimonte 
(Napoli)\footnote{\url{http://www.na.astro.it/tt1/}}. 
The results presented below, out of the 2007
observational campaign, are detailed 
in Calchi~Novati et~al.\cite{novati09}.

\def\figsubcap#1{\par\noindent\centering\footnotesize(#1)}
\begin{figure}[ht]%
\begin{center}
  \parbox{2.1in} {\epsfig{figure=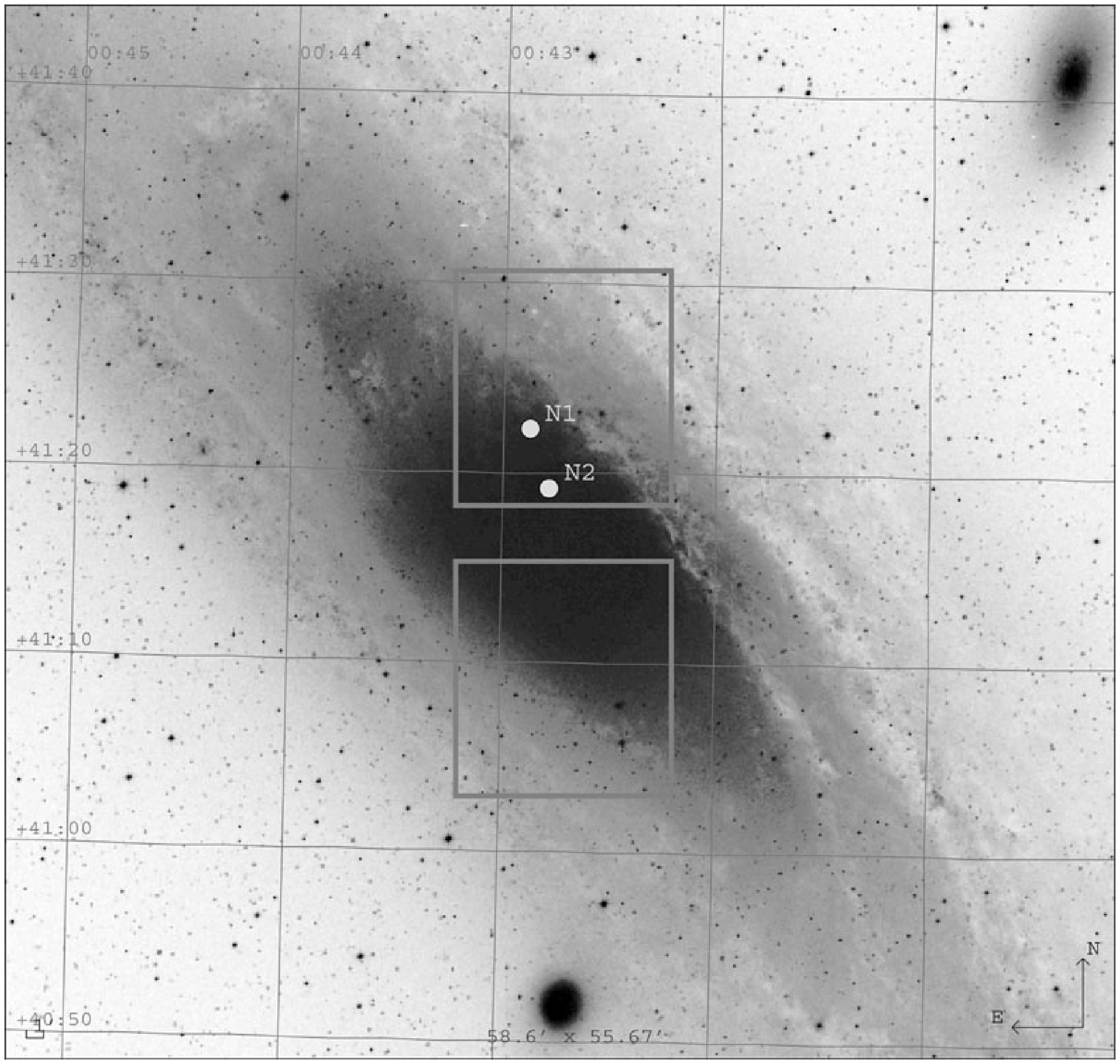,width=2.3in}\figsubcap{a}}
  \hspace*{4pt}
  \parbox{2.1in} {\epsfig{figure=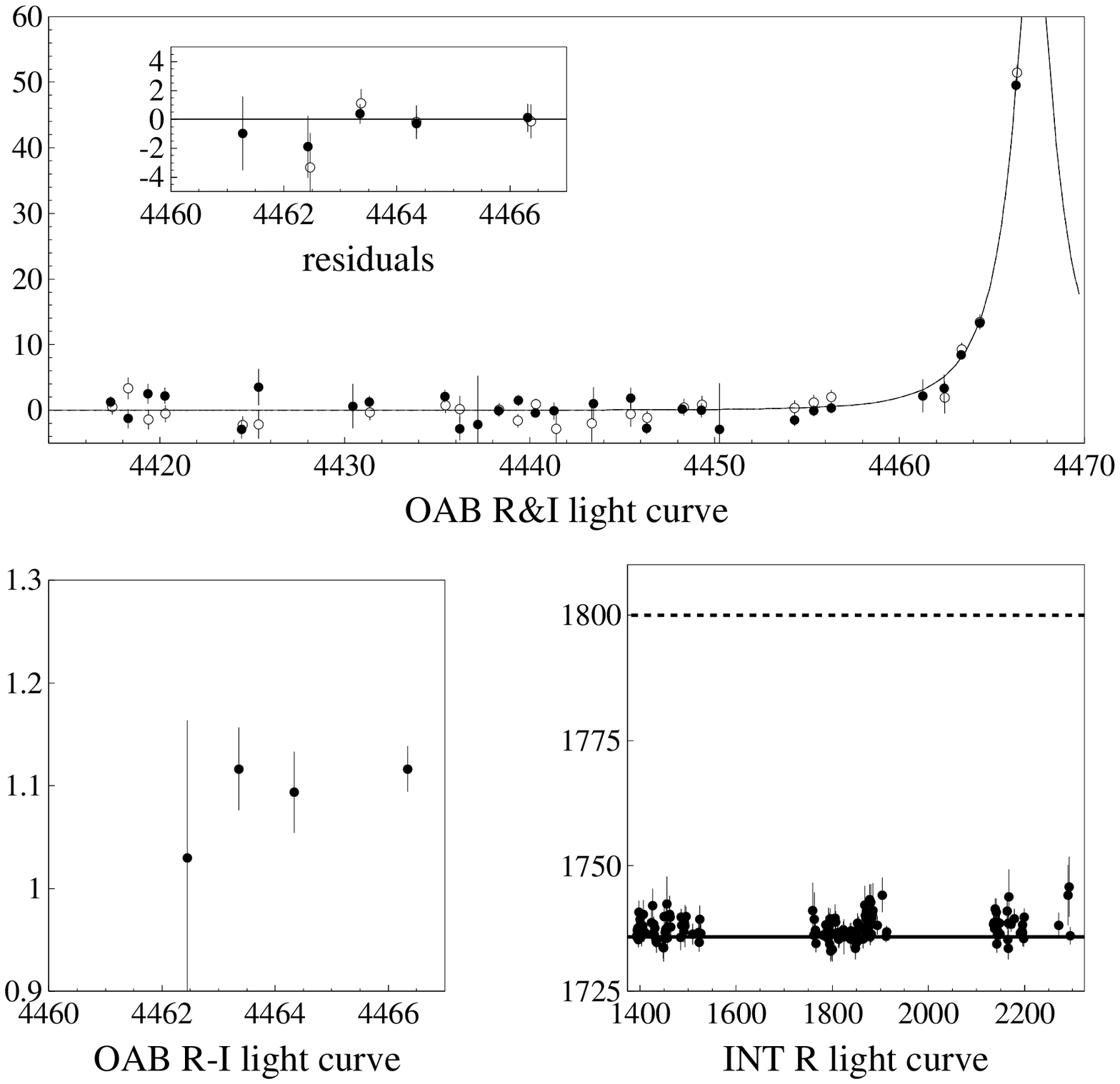,width=2.5in}\figsubcap{b}}
  \caption{(a) Superimposed on a wide field
image of M31 we show the monitored fields of view and the position
of the two candidate microlensing events.
 (b) For the OAB-N2 microlensing
candidate we show the light curve with the best \pacz\ fit, the residual
to the \pacz\ best fit, the achromaticity test light curve 
along the bump and the (flat) extension along the
INT data (the dashed line indicates the flux variation
corresponding to that observed along the OAB data).
The units on the $x$ axes are time in days (JD-2450000.0).
The ordinate axes units are magnitude for the bottom
left panel, and flux in ADU s$^{-1}$ per superpixel for the remaining panels.%
\label{fig1.2}}
\end{center}
\end{figure}

We select microlensing candidates on the basis
of a series of, fully automatised, criteria, among which 
an analysis on the shape
and the sampling of the flux variations, to exclude
intrinsic variable sources,
and an analysis of the PSF at maximum amplification,
to exclude fake events (cosmic rays, bad pixels \ldots).
Furthermore, we limit our search to bright events
with a large enough signal to noise ratio.
As a result we select two microlensing candidate,
OAB-N1 and OAB-N2. Neither of them, however, can
be looked at as a fully convincing candidate.
In particular, we lack data points along the descent
of OAB-N2 (Fig.~1b) so that we can not probe the
expected symmetric shape characteristic of microlensing
events, although the excellent agreement, on the 
raising part, with a \pacz\ shape, together
with its characteristics (duration and brightness at
maximum amplification) are strongly
suggestive of the truly microlensing origin
of this flux variation. In addition to the already
probed flatness of the OAB-N2 light curve extension
along three years of POINT-AGAPE data, the preliminar analysis
of the data of the 2008 and 2009 observational campaigns
at Loiano also confirms the unicity of the OAB-N2 flux variation.

Next step in the analysis, we have to estimate the expected signal.
The number and the characteristics (among which the most relevant,
for a first order analysis,
are the position and the duration) both for M31 self lensing
and MACHO lensing (for which about 1/3 of the
events are expected, for given halo mass
fraction and MACHO mass, to belong to the MW halo). To this purpose
we carry out a full Monte Carlo simulation
of the experiment where we model M31, 
the microlensing amplification
and we take into account the observational set up.
A relevant issue of the above analysis
is the modeling of the M31 luminous components
responsible for the expected self-lensing signal,
to be compared with MACHO lensing:
in particular there is still
a debate about the exact value of the overall mass to be attributed
to the \emph{stellar} lenses, an important parameter
of the model as it is directly proportional
to the number of the expected self-lensing events.
Finally, to properly compare
to the output of the selection pipeline,
we simulate the microlensing events
selected within the Monte Carlo simulation
in the raw data and run from scratch the
full selection process so to reliably
assess the \emph{efficiency} of the pipeline.

Microlensing events towards M31 are expected to be
short duration (usually shorter than 10~days) flux variations,
with, for MACHO lensing, a broader spatial
distribution than for self lensing.
The expected rate we evaluate for our campaign is small.
Somewhat smaller than 1 event for self lensing,
and about the same for MACHO lensing (for istance, 
we expect 2 events for a full halo
of $0.1~\mathrm{M}_\odot$ compact objects).
The detected events are therefore 
compatible with the expected self-lensing rate,
however our available statistics is still too small
to draw definite conclusions on the
dark matter content in the form
of compact halo objects towards M31.
Currently, we are completing the analysis
of the 2008 and 2009 campaigns.

\section{Discussion} 

Pixel lensing searches towards M31
suffer the limitation given by the small expected rate
of events (overall, up to now,
the detection of about 20-30 candidate
events has been reported). This makes particularly troublesome
the disentanglement among self-lensing and MACHO lensing events. 
Analogously to the case of the Galactic center,
the self-lensing signal is an invaluable
tool to probe the stellar
content of M31, and in fact the ANGSTROM collaboration
is focused on this aspect\cite{kerins06}.
On the other hand, for a better understanding of MACHO lensing two main paths may be followed.
First, to increase the statistics of the detected events. 
Besides carrying out long enough observational campaigns,
this means to look for fainter flux variations
(where the risk of contamination
by intrinsic variable sources is however much larger) and, this is essential
to probe MACHO lensing against self lensing,
to map the full area of M31.
In this respect, the PAndromeda project\footnote{S.~Seitz,
talk given at the 13th Microlensing Workshop, 2009, Paris.
This observational programme  is planned  to make use
of the 1.8m PS1 telescope with a huge
field of view of 6.4 sqdeg  
so to cover in a single shot all of the M31 field.
M31 is expected to be monitored with a cadence
of nightly exposures  for about 10 weeks per season.
The first-year campaign has started in fall 2009.}
is expected to represent a real 
``second-generation'' M31
pixel lensing observational programme.
Second, to look for a better light curve characterization and
astrophysical understanding of 
single events. A first example has been
the thorough analysis of the POINT-AGAPE PA-N1 candidate event\cite{point01}.
A second case has been the PA-S3/GL1 candidate event\cite{paulin03,wecapp03},
throughly characterized by the WeCAPP collaboration\cite{arno08},
for which it has been shown, in particular
on the basis of its large amplification at maximum
and on a detailed study of the finite size source effect, that MACHO lensing
is more likely than self lensing. 
As a PLAN collaboration we intend to pursue both these
objectives. First, by completing the analysis
of the 2008 and 2009 observational campaigns. 
Second, by carrying out a more detailed analysis,
making use also of some new data, of the OAB-N2 candidate event\cite{novati10}.
A fundamental aspect of the above mentioned
analyses is the merging of different
data sets. This is relevant both because
it naturally strengthens the microlensing
interpretation of the given flux variation
and because of the increase in the sampling, often
essential for a better characterization of the
event. This is indeed a usual approach for Galactic bulge
light curves, but it clearly shows the path
to be followed also for M31 pixel lensing event analyses.
In this respect, the attempt of the ANGSTROM
collaboration\cite{kerins06} to establish
a real time analysis of M31 pixel lensing 
flux variations, the Angstrom Project Alert System \cite{darnley07} (APAS),
is extremely relevant. This is indeed a first important
step towards a full survey-follow up strategy.  (This is once more a usual approach
for Galactic bulge analyses that would certainly greatly help to improve 
our understanding of the M31 lensing signal.)
Finally, we recall that microlensing is already currently used as a 
powerful tool for the research of extra-solar planets
towards the Galactic bulge\cite{dominik10}.
The coordination among different data sets and observational
campaigns are essential ingredients for the, extremely challenging, purpose
of the research of pixel lensing planet signatures in M31\cite{ingrosso09}.

\bibliographystyle{ws-procs975x65}
\bibliography{biblio}

\end{document}